\documentclass[epjCONF,columns]{svjour}
\usepackage{graphics}
\usepackage{mathptmx} 
\usepackage[latin1]{inputenc}
\session-title{Hadron Collider Physics Symposium 2011}
\begin{document}
\title{Performance of the LHC, ATLAS and CMS in 2011}
\author{Daniel Fournier\inst{1}\fnmsep\thanks{\email{fournier@lal.in2p3.fr}}}
\institute{LAL, Univ Paris-Sud, CNRS/IN2P3, Orsay, France}
\abstract{
The path taken by the LHC team to reach 3.6 10$^{33}$ cm$^{-2}$ s$^{-1}$ instantaneous luminosity, and to deliver 5.6 fb$^{-1}$ per experiment is summarized. The main performances of the two experiments are highlighted, in particular the way they managed to cope with the already high level of ``pile-up''. Selected Standard Model and top physics results are given, and the status of the limits on the Higgs boson search by each experiment is summarized. A brief overview of the search for supersymmetry and exotic phenomena is made at the end.
} 
\maketitle
\section{LHC running in 2011}
\label{sec:1}

\subsection{Machine performance}
\label{subsec:1}
With 5.6 fb$^{-1}$ of good proton-proton data delivered to ATLAS and CMS at a centre of mass energy of 7 TeV, the performance of the LHC is considered by everyone involved as outstanding. The machine also delivered about 1.2 fb$^{-1}$ to LHCb and 5 pb$^{-1}$ to Alice, an integrated luminosity corresponding to the running conditions chosen by these experiments.

The main parameters of the machine, at the end of the running period where the instantaneous luminosity was\linebreak highest, are listed in table 1, together with the running parameters at the end of 2010, and in comparison with the nominal parameters at 14 TeV in the centre of mass \cite{Ref1}. The beam crossing angle at the collision points 1 and 5 where ATLAS (resp CMS) are installed was of 120 microradians, sufficient to limit long range beam-beam effects for a bunch spacing of 50 ns. 

\begin{table}[h]
\vspace*{-0,4cm}
\caption{Parameters of LHC exploitation, at the end of 2010, at the end of 2011, and design parameters at 14 TeV in the centre of mass.}
\label{tab:1}       %
\begin{tabular}{llll}
\hline\noalign{\smallskip}
Parameter & 2010 & 2011 & Nominal  \\
\noalign{\smallskip}\hline\noalign{\smallskip}
N (10$^{11}$ p/b) & 1.2 & 1.5 & 1.15\\
k (n$_{bunches}$) & 368 & 1380 & 2808 \\
Bunch spacing (ns) & 150 & 50 & 25 \\
$\epsilon (\mu$m rad) & 2.4-4 & 1.9-2.3 & 3.75\\
$\beta^*$ (m) & 3.5 & 1 & 0.55\\
L (cm$^{-2}$ s$^{-1}$)  & 2 10$^{32}$ & 3.6 10 $^{33}$ & 10$^{34}$\\
Energy (MJ) stored & 28 & 110 & 360\\
\noalign{\smallskip}\hline
\end{tabular}
\vspace*{-0,2cm}
\end{table}

The path from the 2010 parameters to those obtained at the end of this year, which represent an improvement by a factor 20, was made in successive steps:
\vspace*{0,1cm}

\noindent - restart of the machine with trains of bunches separated by 150 ns

\noindent - beam scrubbing at injection energy and high intensity to ``clean-up'' the sections prone to electron-cloud effects

\noindent - running with short trains of bunches separated by 50 ns

\noindent - intensity ramp-up with up to 1380 bunches per beam (maximum possible with 50ns spacing)

\noindent - emittance reduction

\noindent -  $\beta^*$ reduction from 3.5m down to 1m

As of  early September a peak luminosity of about 3.6 10$^{33}$ at the beginning of a fill was reached, and reproduced until the end of the period (end October), allowing to record about
half of the total integrated luminosity of the year during the last two months (Fig. 1).

\begin{figure}[h]
\vspace*{-0,2cm}
\centering
\resizebox{0.75\columnwidth}{!}{%
  \includegraphics{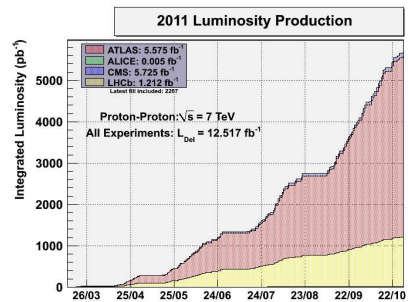} }
\vspace*{-0,2cm}
\caption{Luminosity integrated by experiments in 2011.}
\label{fig:1}
\end{figure}

These excellent performances were obtained despite a number of effects which limited the machine availability and the actual length of most of the fills. Some of these effects \cite{Ref2} are briefly discussed below:
\vspace*{0,2cm}

\noindent - Radiation induced failures of Electronics (SEU). The cryogenics and the machine protection systems suffered from electronics failures due to radiation effects, whose rate was shown to be nearly proportional to the instantaneous luminosity. A number of actions were taken during the year to mitigate the incidence of such effects (relocation or shielding of critical electronics, improvement of redundancy,...).

\noindent - Beam dumps triggered by high losses (UFO). Sudden increase of beam losses during stable conditions have been tentatively associated to ``falling objects''. By far not all these losses  trigger a beam dump. While during injection most UFOs happen in the vicinity of kicker magnets, losses during stable beam are more or less uniformly distributed around the machine. It was observed that the rate of UFOs tend to decrease during long periods of reproducible conditions. 

\noindent - Vacuum pressure increases. Very significant pressure increases (factors 100 or more) appeared when trains of\linebreak bunches separated by 50 ns were injected in the machine. They were principally located around the collision points, at transition places between cold and warm sections of the machine. Dedicated periods at high intensity and injection energy (beam scrubbing) allowed to ``clean up'' the\linebreak critical places, with pressure reductions of typically an order of magnitude after 15 hours. Wrapping simple coils around these places furthermore reduced the effect, confirming that the source was dominantly ``electron-clouds''.

\noindent - Heating of beam elements at various places (beam screen, kickers, collimators, and -recently discovered- damaged RF-fingers), associated to the electromagnetic field accompanying the bunches, was observed and forced to limit the bunch charge in several occasions.

\noindent - RF beam loading and beam instabilities leading to emittance blow-ups were also observed at several occasions.

As a global consequence of these various effects, the duration of beam fills was in average 6 hours, with large variations, while the optimum would have been around 12 hours. Another measurement of the impact of the above limitations, is through the machine availability, which was about 50\% during the ``physics periods''. In turn about half of this time corresponded to ``stable beam condition'' which covered thus  23\% of  ``physics'' time (Fig. 2).

\begin{figure}[h]
\centering
\resizebox{0.75\columnwidth}{!}{%
 \includegraphics{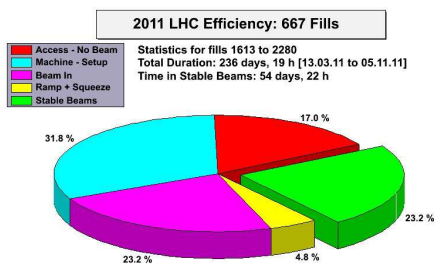} }
\vspace*{-0,2cm}
\caption{Pie chart of machine availability.}
\label{fig:2}
\vspace*{-0,4cm}
\end{figure}

\subsection{Luminosity measurement and adjustments}
\label{subsec:2}

Knowing the luminosity with precision is an important asset for many, if not all physics measurements. In 2010 the Van der Meer method was applied to measure the luminosity with an accuracy of 3.4\% in ATLAS \cite{Ref3} and 4.0\% in CMS \cite{Ref4}. In 2011 this was repeated twice. The method applied at the LHC goes as follows:
    \vspace*{0,1cm}

\noindent - the machine is run with a small number of bunches

\noindent -  the beam current associated to each bunch is measured with a ``beam current transformer'' (giving n$_1$ and n$_2$)

\noindent - the size of the luminous region is measured by studying the counting rate as a function of beam separation, sequentially in the horizontal and vertical directions, giving ${\textstyle\sum_x}$ and ${\textstyle\sum_y}$. This measurement is made using luminosity monitors, which are also run during high luminosity data taking.
\vspace*{0,1cm}
L is obtained as ($f_r$ is the rotation frequency of one bunch around the ring):

\begin{equation}
L = k.f_r.n_1.n_2/(2\pi{\textstyle\sum\nolimits_x\sum\nolimits_y})
\end{equation}

\noindent -  integrated luminosity of long physics periods, is then obtained summing up counts from luminosity monitors, suitably corrected for various effects (\cite{Ref3}, \cite{Ref4}), and normalized to the VDM scan period.
\vspace*{0,1cm}

An example of Van der Meer scan, taken from ATLAS is shown in Fig. 3. 

\begin{figure}[h]
\centering
\resizebox{0.75\columnwidth}{!}{%
  \includegraphics{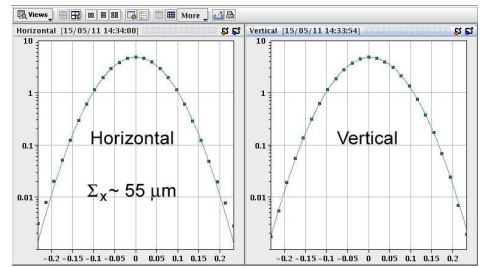} }
\caption{Example of counting rate during a Van der Meer scan.}
\label{fig:3}
\end{figure}

The size of the luminous region is related to the counting rate R by:

\begin{equation}
             {\textstyle\sum\nolimits_x} = \frac{1}{\sqrt{2\pi}} \frac{\int R_x (\delta) d\delta}{R_x (0)}
\end{equation}

The accuracy of the method is limited by the bunch charge measurement accuracy, and by non linearities in the luminosity monitors. The precision obtained for the first part of the 2011 data (until $\beta^*$ was lowered to 1m) is 3.7\% for ATLAS and 4.5\% for CMS.

The reliability of running the machine with beams\linebreak separated in one direction serves another important purpose. It allows to run the machine with ``luminosity levelling'' in LHCb, around 2 to 3 10$^{32}$ cm$^{-2}$ s$^{-1}$ as requested by this experiment. Fig. 4 shows an example, for a given fill of how well the instantaneous luminosity is kept constant in this experiment, while it smoothly decreases in ATLAS and CMS because of  proton losses and emittance increase.

\begin{figure}[h]
\centering
\resizebox{0.75\columnwidth}{!}{%
  \includegraphics{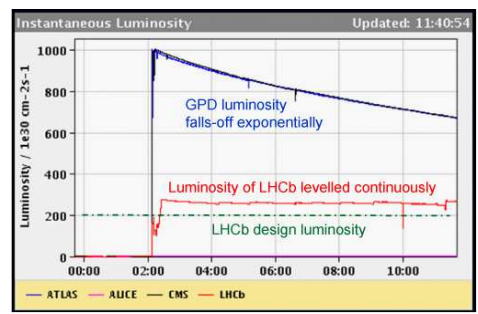} }
\caption{Example of a fill with luminosity levelling in LHCb.}
\label{fig:4}
\end{figure}

Finally the low luminosity in Alice is provided by colliding in point 2 a few dedicated bunches added to the trains.

As a consequence of the complex bunch structure in each ring, a few ``unpaired bunches'' cross the nominal collision points at a time when there is no bunch coming from the other direction, thus without producing any collision. The data corresponding to these BCIDs (bunch crossing identifiers) are however recorded and used to evaluate the ``non-collision'' background, associated for the largest part to beam losses in the arcs. Secondaries produced when these lost protons interact in the magnet yokes, or muons from the subsequent decay of the secondaries, eventually reach the experiments and activate some low level triggers. Analysing these data shows \cite{Ref5} that, as anticipated, they mostly consist of energy deposits located in the (horizontal) machine bending plane. Their rate is very low, confirming the very high efficiency of the collimation process in the machine. This background nevertheless needs to be rejected for some analyses looking for rare processes with rather weak signatures (jets and missing transverse momentum E$_T^{miss}$ for example).

\subsection{Pile-up }
\label{subsec:3}

As will be discussed in the next sections, the occurrence of several independent, inelastic, proton-proton collisions during one  bunch crossing constitutes a noise, usually called pile-up noise, which degrades to some extent the performance of the reconstruction of some of the ``objects'' used for physics analysis. It is thus important to be able to characterize the pile-up conditions on an event by event basis. One essential variable in this respect is the number of primary vertices reconstructed per crossing, called N$_{VX}$ in the following. Since the response of several detectors (in particular the calorimeters) extends over more than the time interval between two successive crossings, it is also important to have a view of the pile-up conditions in the surrounding bunches (so called ``out of time'' pile-up as opposed to ``in-time'' pile-up described by N$_{VX}$). The variable used for describing the pile-up conditions overall is $\mu$, the number of interactions per crossing. $\mu$ is calculated from the suitably normalized instantaneous count of luminosity monitors (L) and the inelastic cross-section as:

\begin{equation}
\mu = L.\sigma_{inel}/k.fr
\end{equation}

Summing up the entries  over a data taking period, one obtains the histogram of the ``luminosity weighted mean number of interactions per crossing'' as shown in Fig. 5 for ATLAS. The dispersion reflects the variation of charges between bunches, the decrease of instantaneous lumino\-sity during each fill, and the difference of initial conditions from fill to fill. In particular the change between $\beta^*$ values is striking, with $<\mu>$ going from 6.3 for $\beta^*$=1.5m, to 11.6 for $\beta^*$=1.0m. In average N$_{VX}$ and $<\mu>$ are roughly proportional, with 
$<$N$_{VX}$$>$$\sim$0.6$<\mu>$, the smaller number of reconstructed vertices resulting from the limited detector acceptance, and the minimal conditions set for a valid reconstructed vertex (typically at least 3 tracks of at least 0.5 GeV/c transverse momentum).

\begin{figure}[h]
\centering
\resizebox{0.75\columnwidth}{!}{%
  \includegraphics{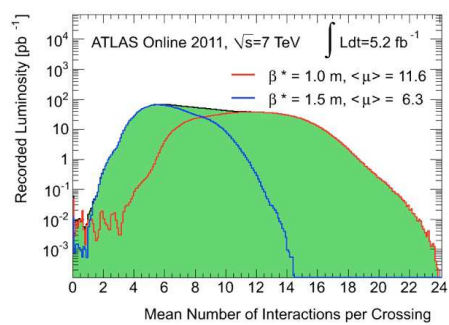} }
\caption{Luminosity weighted mean number of interactions per crossing.}
\label{fig:5}
\end{figure}

\subsection{Special runs}
\label{subsec:4}

On top of regular ``physics runs'' in nominal conditions, some runs were taken in peculiar conditions to address specific questions:
\vspace*{0,1cm}

\noindent - Data were taken with short trains of bunches separated by 25ns in order to assess the trigger, data taking, and reconstruction performance in the future ``nominal'' conditions.

\noindent - Fat bunches, of close to twice the nominal charge were also collided in order to assess the pile-up effects with\linebreak $<\mu>$ reaching values of up to 40.

\noindent - Several days in Sept 2011 were dedicated to data taking with high $\beta^*$ settings (90m) in view of measuring elastic scattering with dedicated forward detectors (ATLAS/ALFA and TOTEM).  These detectors positioned at 240m on each side of the collision points 1 and 5, are located in ``Roman pots'' and allow to trigger and record elastically scattered protons with transverse momenta up to about 1.5 GeV/c \cite{Ref6}.The beam conditions were clean enough that the detectors could be brought down to 6$\sigma$ from the beams for data taking. More than 1 million elastic events were recorded by each experiment. Higher $\beta^*$ are planned in 2012 (up to 1km) in view of reaching the Coulomb interference region.

\section{ATLAS and CMS status }
\label{sec:2}

The structure of each of the two experiments is known worldwide already. 
ATLAS \cite{Ref7} features an air core toroid, high granularity ``accordion'' lead-liquid argon electromagnetic calorimetry, complemented by Copper/liquid argon and tungsten liquid argon hadronic calorimeters in the forward direction,  and iron- scintillating tiles in the central region. The central tracking, inside a 2T magnetic field, uses Si-pixels and Si-strips in the inner part, and straw tubes at larger radius.

In CMS \cite{Ref8} the large solenoid of 4T magnetic field contains the inner detector with pixel and Si-strips, embedded in a PbWO$_4$ crystal calorimeter, followed by a brass-scintillating tile hadronic calorimeter. The forward calorimeters are recessed at 8m from the collision point and use a steel-quartz fiber (Cerenkov) sampling technique. Both experiments have a sophisticated 3-level trigger system, making a large use of lepton (electrons, muons, taus), high energy jet and  E$_T^{miss}$ signatures.

Up to now, the detectors have not suffered significant radiation damage. One observes however that the leakage current in Si-pixel sensors is increasing, in particular in CMS where, for technical reasons, the pixel detector is for the time being operated at warm temperature. The type inversion is expected to happen some time in 2012.

During the past year about 4 10$^{14}$ events were processed by the trigger system of each experiment, sollicitated  at the 20 MHz bunch collision frequency. At the other end, about  300 events/s, in average, were written to permanent storage by each experiment. It is instructive to look at tables 2 and 3 which show for each experiment, at 3 10$^{33}$ peak luminosity, how the bandwidth is split between the main trigger/DAQ channels, keeping in mind that the full trigger menu has about 350 lines in ATLAS, and even more in CMS. A good fraction of them are used to monitor the trigger itself by accepting, at each level, a prescaled fraction of the rejected events. Among the main signatures are the presence of one identified lepton. One can observe that the lepton thresholds are somewhat lower in ATLAS, thus taking  in comparison a larger fraction of the bandwidth. Given the steep slope of the lepton (and jets) transverse momenta, it is extremely important to have sharp turn-on thresholds to reduce the rate of unwanted signals. As an example, Fig. 6 shows the turn-on curve for the combined muon and hadronic tau trigger, as a function of the offline tau-jet transverse momentum. 

\begin{figure}[h]
\centering
\resizebox{0.75\columnwidth}{!}{%
  \includegraphics{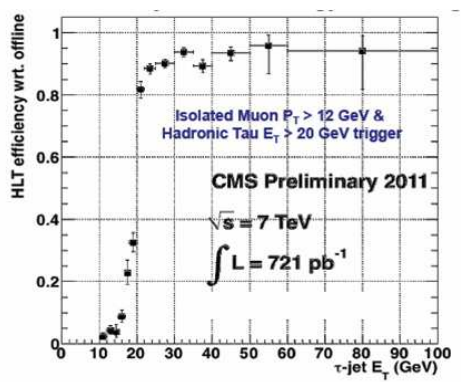} }
\caption{Turn-on curve for the combined muon (12 GeV) and hadronic tau trigger, in CMS.}
\label{fig:6}
\end{figure}

At the storage level, the data volume is larger for ATLAS ($\sim$2 Mbyte/event) than for CMS ($\sim$ 0.5 Mbyte/event) where zero suppression in the calorimeter is applied at the data acquisition level.

\begin{table}[h]
\caption{ATLAS trigger threshold and rates at the LVL1 and after final selection.}
\label{tab:2}       %
\resizebox{1.0\columnwidth}{!}{%
  \includegraphics{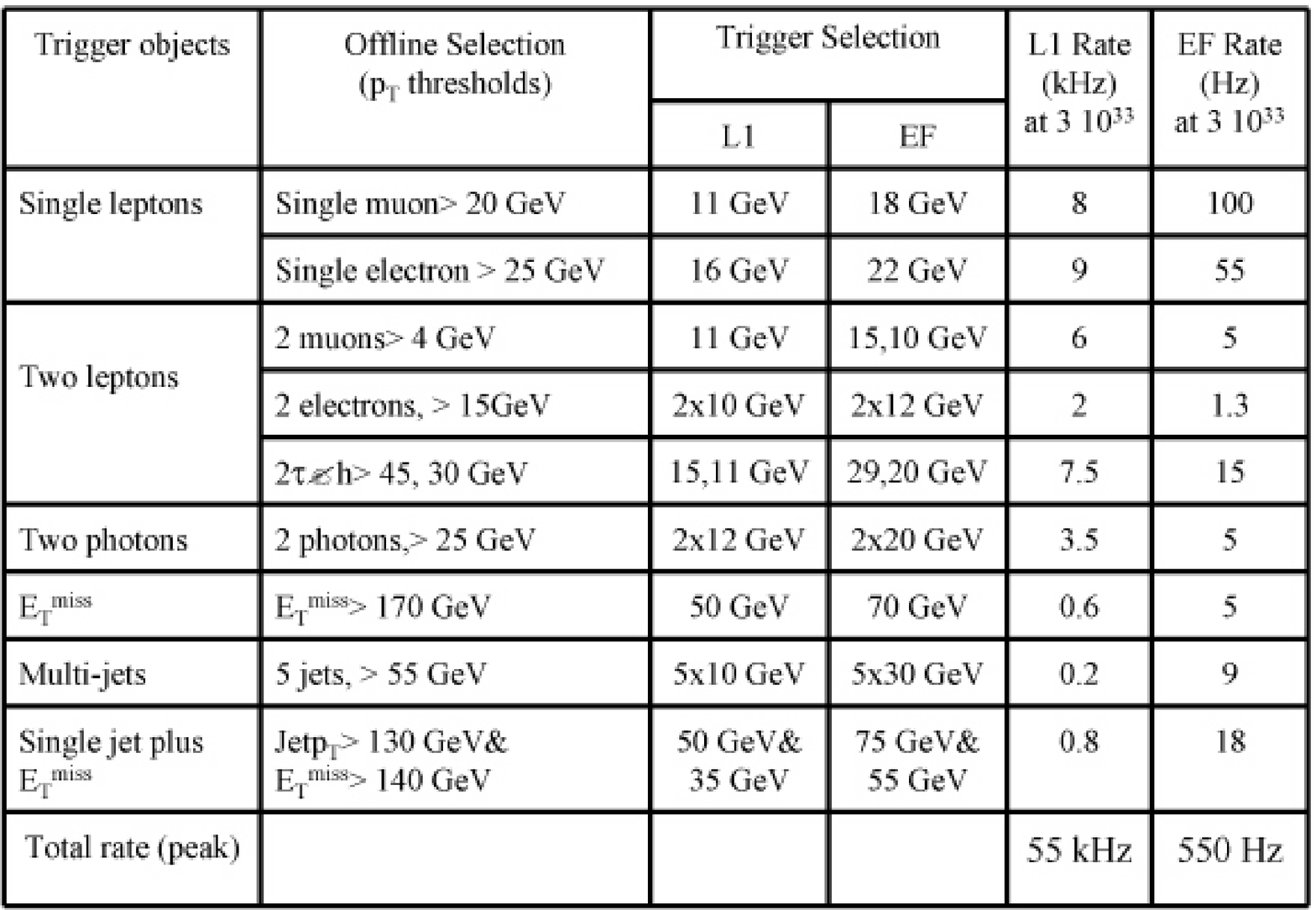} }

\end{table}

\begin{table}[h]
\caption{CMS trigger threshold and rates at the LVL1 and after final selection.}
\label{tab:3}       %
\resizebox{1.0\columnwidth}{!}{%
  \includegraphics{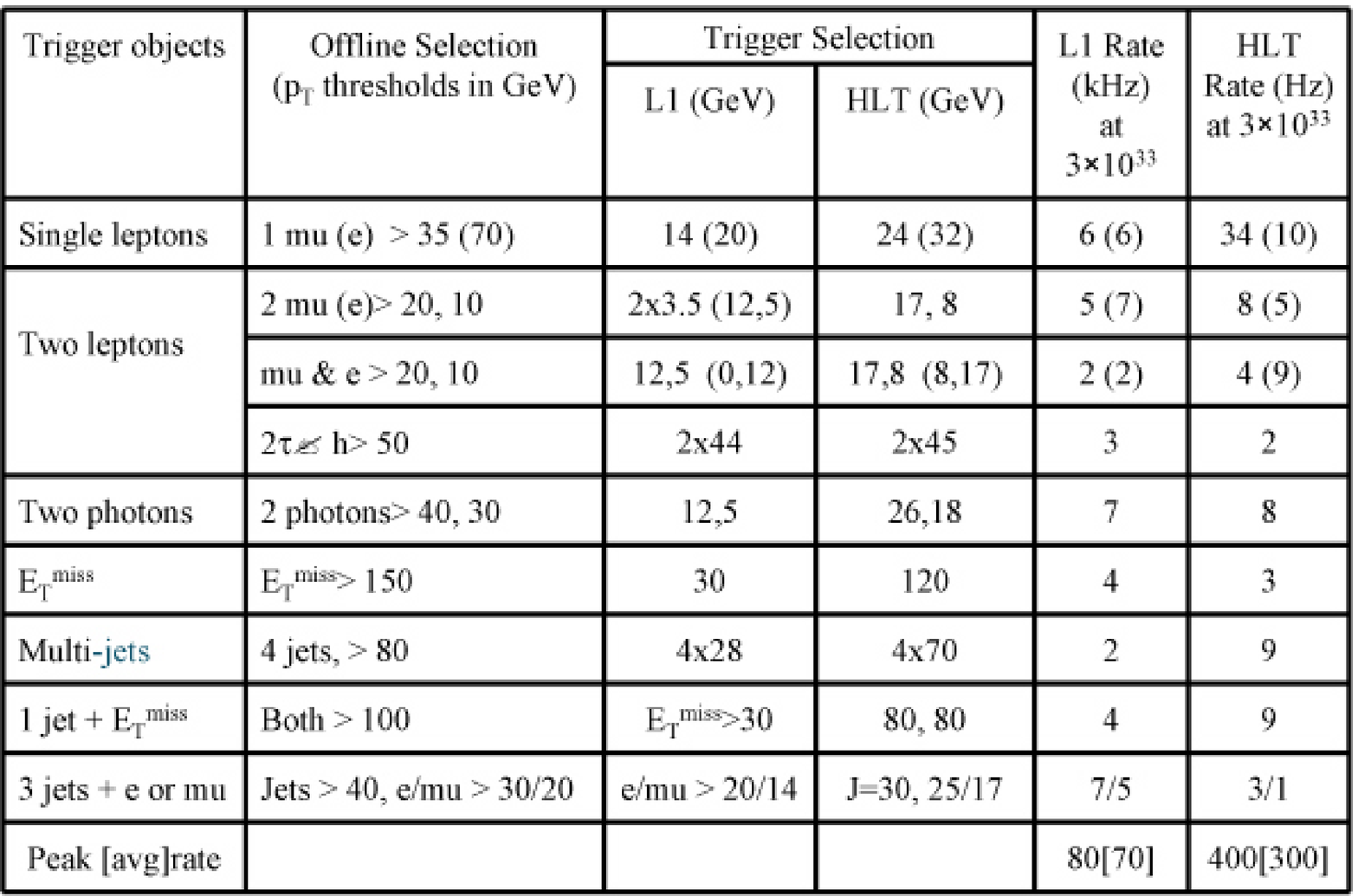} }
\end{table}

The combined efficiency of data acquisition and data quality selection in each experiment is at a high level, allowing to find close to 90\% of the delivered luminosity in the physics plots.

The fraction of channels alive is also on a high standard, being for example 99\% or more for each of the two electromagnetic calorimeters, and about 97\% for each of the two pixel systems.

\section{Physics objects and selection of SM results}
\label{sec:3}

The new feature of 2011 data was the high level of pile-up, with $<\mu>$ reaching  up to about 20 at the end of the period. While the pile-up ``noise'' is not expected to affect very high energy jets, nor lepton reconstruction, nor even b-tagging, it is on the other hand expected to affect seriously objects of large size and low/medium transverse energy, and therefore in particular: 

\begin{itemize}
\item low energy jets
\item E$_T^{miss}$
\item isolation of leptons and photons.
\end{itemize}

\subsection{Jets and QCD}
\label{subsec:3-1}

In the high transverse momentum (p$_T$) range, the critical quantities for jets are the energy scale and the linearity. In ATLAS jets are reconstructed with the anti-kT algorithm, with a size parameter R=0.6 (R=0.4 is also used for complex final states). Jets are built as vectorial sums of clusters of calorimeter cells, corrected for hadronic to electromagnetic response, and dead material losses. In situ methods are used to check the p$_T$ scale up to 1 TeV or above: photon-jet balance, multi-jet balance, track-jets,..

In 2010 the energy scale systematic uncertainty was 2.5\% in a wide kinematic range. The large data set of 2011 may allow to improve this uncertainty.
The ATLAS inclusive p$_T$ spectrum, with 1.9 fb$^{-1}$ integrated luminosity, also including 2010 data on the lower p$_T$ part, is shown in Fig. 7. The data is well reproduced by a PYTHIA simulation in which the PDF are corrected for NLO effects\cite{Ref9}. Jets with p$_T$ up to 1.9 TeV have been observed!

\begin{figure}[h]
\centering
\resizebox{0.75\columnwidth}{!}{%
  \includegraphics{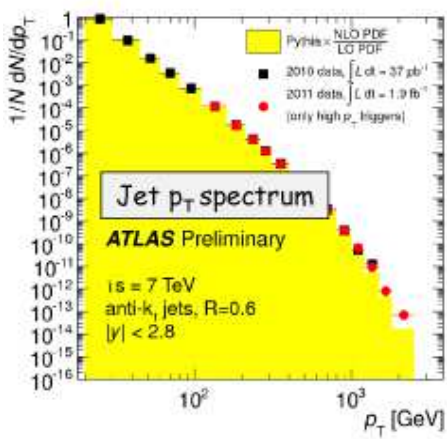} }
\caption{Inclusive jet p$_T$ distribution in ATLAS.}
\label{fig:7}
\end{figure}

In CMS jets are reconstructed from ``Particle flow'' clusters, using the anti-kT algorithm with R=0.5. A certain level of cluster merging (``jet grooming'') is made before going to physics distributions. As an illustration, the invariant mass spectrum of the two jets of higher p$_T$ is shown in Fig. 8, compared to PYHTIA simulation (with CTEQ6L1) scaled up by a factor 1.33. The agreement is very good, and allows to rule out excited quarks mith masses smaller than 2.49 TeV (2.68 expected limit), Axigluons with masses lower than 2.47 TeV (2.66 expected), and W' with masses smaller than 1.51 TeV (1.40 expected).

\begin{figure}[h]
\centering
\resizebox{0.75\columnwidth}{!}{%
  \includegraphics{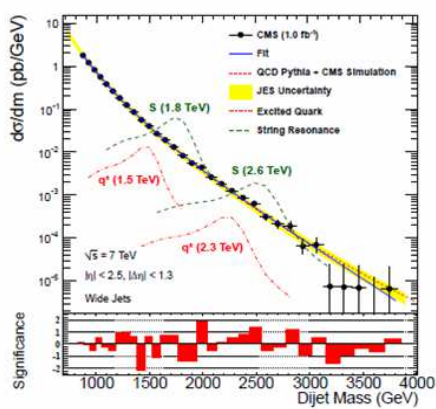} }
\caption{Invariant mass of leading jets in CMS.}
\label{fig:8}
\end{figure}

\subsection{Impact of pile-up on sensitive quantities}
\label{subsec:3-2}

\noindent
- Jet energy resolution. The p$_T$ balance between photons, insensitive to pile-up given their small size, and jets allows to control the jet energy scale and resolution. In ATLAS it was found, in the first part of the data ($<\mu>$=6), that the jet energy resolution is worsened by 10\% in the lowest p$_T$ range (30 GeV). In CMS the effects of ``in time pile-up'' and of ``out of time pile-up'' were separated showing that (for N$_{VX}$=8) the latter contributes about 5 GeV rms to the component perpendicular to the photon direction while the former contributes about 2 times more. These figures correspond to the present status, without any correction to mitigate the observed worsening.

\noindent 
- Lepton and photon isolation. In order to reduce the background of fake muons coming from heavy quark decays inside jets, an isolation cut is often applied. It consists either of a track isolation cut (sum of transverse momenta of tracks falling inside a cone of size $\Delta$R around the muon), or a transverse energy cut (sum of all calorimeter transverse energies) or a combination of both. Fig. 9 shows the spectrum of the calorimeter transverse energy in a cone of $\Delta$R=0.4 around muon tracks from Z$^0$ decays in ATLAS, for two pile-up conditions corresponding to N$_{VX}$=4 and N$_{VX}$=8. With higher pile-up one observes a broadening of the distribution, and a shift of the mean value. The latter can be subtracted from an estimate of the ``ambient'' pile-up level, but the broadening of course will stay. The width approximately doubles when N$_{VX}$ goes from 1 to 12, meaning that  the underlying event of Z$^0$ production is more busy than for a random event.
\begin{figure}[h]
\centering
\resizebox{0.75\columnwidth}{!}{%
  \includegraphics{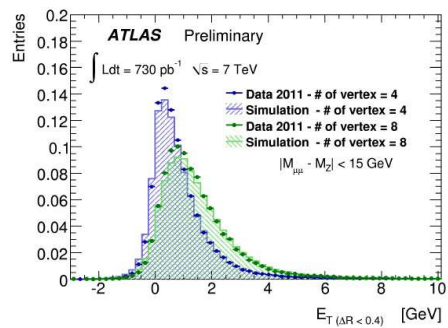} }
\caption{Muon isolation in ATLAS.}
\label{fig:9}
\end{figure}

\noindent
-  E$_T^{miss}$.
Obtained as the negative vector sum of the transverse momentum of all ``objects'' in an event, the E$_T^{miss}$ is a priori quite sensitive to pile-up. This dependence is limited if one retains only jets above a minimum p$_T$. As an example of performance, for a rather complex final state, Fig. 10 shows E$_T^{miss}$ in  CMS (first part of 2011 data) for events with 2 leptons of opposite charge and identical flavor, in the Z$^0$ mass range, plus 2 jets. One observes that the peak of  Z+ jets events remains rather narrow, and that the tail at high E$_T^{miss}$ (greater than about 70 GeV) is dominated by physical processes, essentially $t\bar t$ pairs.

\begin{figure}[h]
\centering
\resizebox{0.85\columnwidth}{!}{%
  \includegraphics{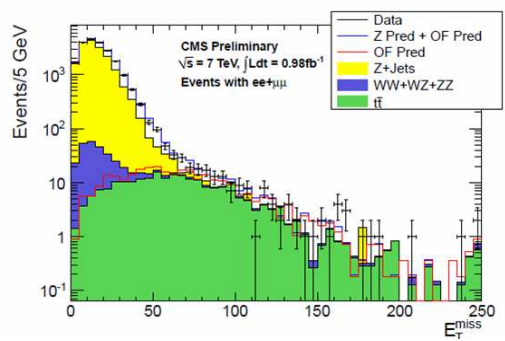} }
\caption{E$_T^{miss}$ distribution for Z+ 2 jets events in CMS.}
\label{fig:10}
\vspace*{-0,8cm}
\end{figure}

\subsection{W\&Z Physics}
\label{subsec:3-3}

This physics is entirely done with the leptonic decay modes. The full 2011 data set  represents in each experiment about 3 millions of Z decays in electron pairs or muon pairs, and 10 times more W decays in electron-neutrino or muon-neutrino. The decays to $\tau$ come on top of this, with significantly smaller statistics due to the reduced trigger and reconstruction efficiencies. Given the low level of background under the peak, Z decays are used to establish with the ``tag-and-probe'' method the trigger and reconstruction efficiencies in data and in Monte-Carlo simulations. They are also used to set the electron and the muon energy scales, and where necessary to improve the energy (electrons) or the momentum (muons) reconstruction.

Early physics results were obtained with the 2010 data, illustrated by Fig. 11 which shows the W and Z fiducial cross-sections in ATLAS \cite{Ref10}, compared to calculations at the NNLO using different PDF sets. Also particularly sensitive to the PDFs is the charge asymmetry of  leptonic W decays. The variation of this asymmetry with the lepton pseudo rapidity (see Fig. 12) reflects combined effects of PDFs, W polarization and V-A decay of the W. The sign inversion around $\vert \eta\vert$ =3.0, which falls into the acceptance of the LHCb experiment is nicely reproduced by simulations. Larger data sets, as now available, will certainly improve the knowledge of PDFs.

\begin{figure}[h]
\centering
\resizebox{0.75\columnwidth}{!}{%
  \includegraphics{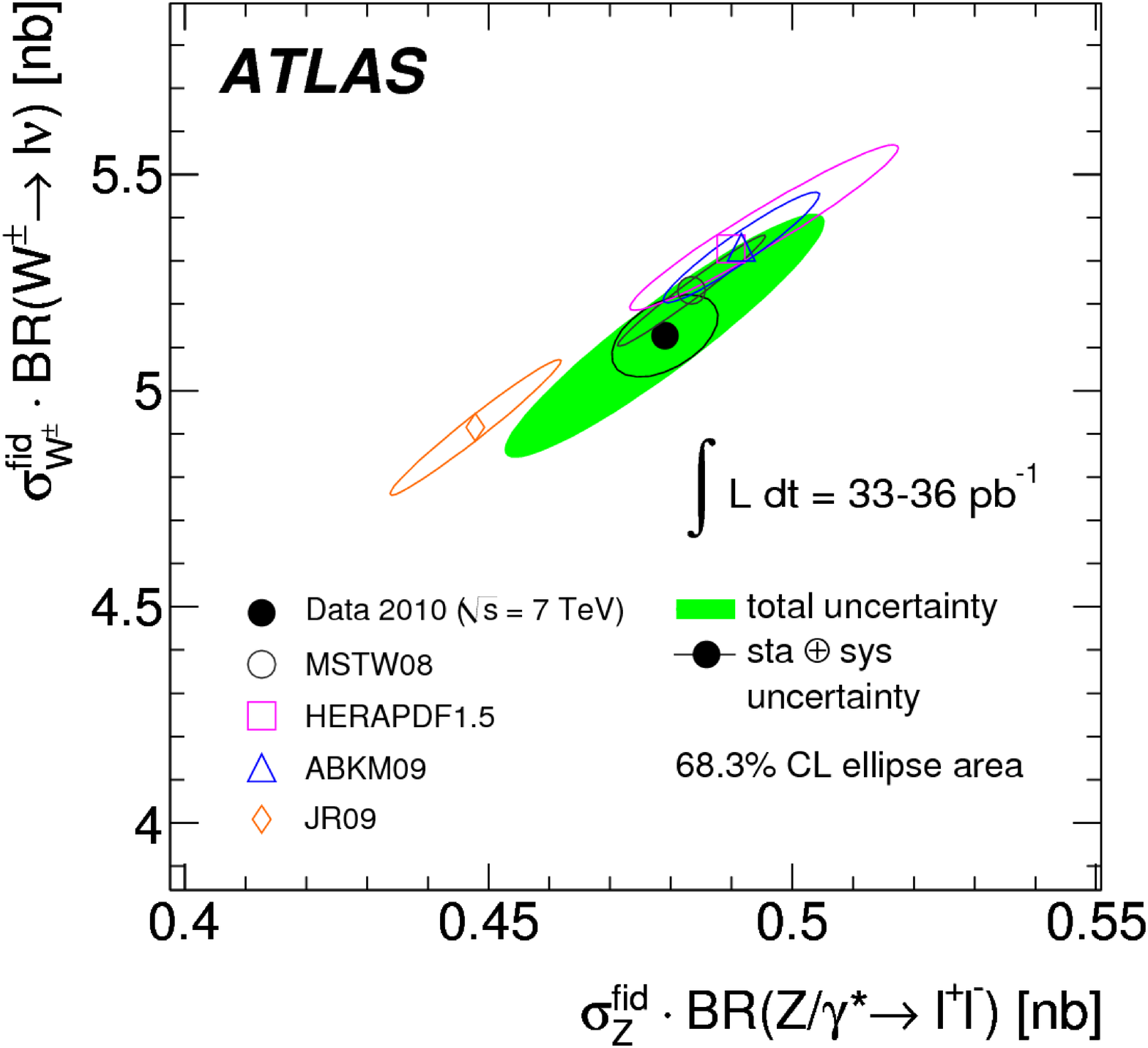} }
\caption{W and Z fiducial cross-sections in ATLAS, compared to NNLO simulations.}
\label{fig:11}
\end{figure}

\begin{figure}[h]
\vspace*{-0,3cm}
\centering
\resizebox{0.75\columnwidth}{!}{%
  \includegraphics{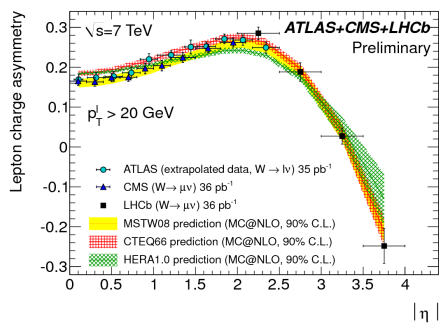} }
\caption{Leptonic charge asymmetry in W decays (ATLAS, CMS, LHCb) compared
to NLO simulations.}
\label{fig:12}
\end{figure}

Z production is the best possible place to assess the efficiency and the accuracy of $\tau$ decays reconstruction. The trigger efficiency is the first problem to be overcome, given the large fraction of transverse momentum taken by the neutrinos in the final state. To reach low enough thresholds, double conditions are required, like illustrated for example in Fig. 6 for the $\mu$-had final state. The hadronic ``$\tau$-jets'' are separated from electron showers and from jets by a combination of criteria on charged tracks (1 or 3), on shower shapes in the electromagnetic and hadronic calorimeters, and on isolation. Finally to isolate Z decays, a minimum E$_T^{miss}$ is required. Fig. 13 shows the spectrum obtained by CMS \cite{Ref11} in the muon-hadron channel, still with 2010 data. The cleanliness of the signal, together with a Z cross-section in the $\tau\tau$ mode ($\sigma$.BR=1.0 $\pm$ 0.05 (stat) $\pm$ 0.08(syst) $\pm$ 0.04(lumi) nb) which matches the other leptonic decay modes demonstrate that $\tau$ decays are mastered, which is an important asset for many physics objectives. Indeed a worsening of missing momentum resolution will somewhat degrade the situation. The $\tau$ performances in ATLAS are similar to CMS.

\begin{figure}[h]
\vspace*{-0,3cm}\centering
\resizebox{0.75\columnwidth}{!}{%
  \includegraphics{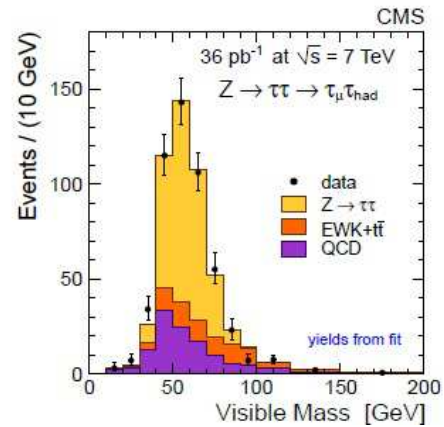} }
\caption{Z decays reconstructed  in the lepton-hadron decay mode (CMS).}
\label{fig:13}
\vspace*{-0,5cm}\end{figure}

\subsection{B-tagging and top physics}
\label{subsec:3-4}

Efficient B-tagging is a key to top physics, to some Higgs channels,... Already commissioned with 2010 data, ``advanced tagging methods'' were validated with the 2011 data set. Among them ATLAS uses a combination of the track impact parameter in 3D (IP3D) and of a fit of secondary vertices (SV1). At 60\% efficiency, this combined approach has a rejection 4 times larger than the early ``SV0'' algorithm \cite{Ref12}. Fig. 14 illustrates this performance by showing the fraction of jets satisfying the  b-tagging cut at 60\% efficiency, compared to Monte-Carlo simulation. The agreement is satisfactory, and shows that, around 100 GeV, 50\% of the events passing the cut are genuine b-jets while 60\% of the remaining ones are actually charmed jets.

\begin{figure}[h]
\centering
\resizebox{0.75\columnwidth}{!}{%
  \includegraphics{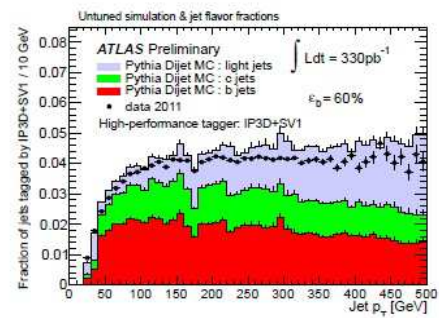} }
\caption{Fraction of b-jets tagged by ``IP3D+SV1'' in ATLAS.}
\label{fig:14}
\end{figure}

The production of $t \bar t$ pairs was already measured  with 2010 data, both in the single lepton and in the dilepton modes, with and without b-tagging. With the first 0.7 fb$^{-1}$ of 2011 data, ATLAS measured  the cross section using both modes \cite{Ref13}.

\begin{equation}
\sigma^{t\bar t}_{ATLAS} = 179.0 \pm 3.9\ (stat) \pm 9\ (syst) \pm 6.6\ (lumi)\ pb
\end{equation}

Top physics is also an important part of the CMS program. The experiment measured the cross-section \cite{Ref14}

\begin{equation}
\sigma^{t\bar t}_{CMS} = 166 \pm 2.2\ (stat) \pm 11\ (syst) \pm 8\ (lumi)\ pb
\end{equation}

These values, about 20 times larger than at the Tevatron, are to be compared to the calculated NNLO cross-section of 164.6 $\pm$ 13 pb.

The top mass measurements in ATLAS and CMS are affected by systematic uncertainties (final state radiation, b-jet energy scale) which are still larger than at the Tevatron. However, in the single lepton channel, comparing positive an negative muon decays, CMS measured with 1.09 fb$^{-1}$ of 2011 data, the top-antitop mass difference with a reduced systematic uncertainty \cite{Ref15}:

\begin{equation}
\Delta m=-1.20 \pm 1.21\ (stat) \pm 0.47\ (syst)\ GeV
\end{equation}

\noindent
which is the most precise value so far.

Observing ``single top'' production at the Tevatron was a real challenge for several years. Thanks to the higher centre of mass energy, both experiments at the LHC reported single top observation with  2010 data  already. More accurate results with the first part of the 2011 data were already made public by ATLAS \cite{Ref16}. The ``t-channel'' analysis (exchange of a W boson in the t channel) requires 1 lepton, 1b-jet and 1 or 2 more jets, and  E$_T^{miss}$ in the final state. A clear signal was observed (see Fig. 15), and the cross-section was measured to be \cite{Ref16}:

\begin{equation}
\sigma^t_{ATLAS}  = 90\pm 9\ (stat) ^{+31}_{-20}\ (syst)\ pb
\end{equation}

\noindent
to be compared with the ``approximate NNLO'' prediction of 64.6 $\pm$ 3 pb.

\begin{figure}[h]
\vspace*{-0,2cm}\centering
\resizebox{0.75\columnwidth}{!}{%
  \includegraphics{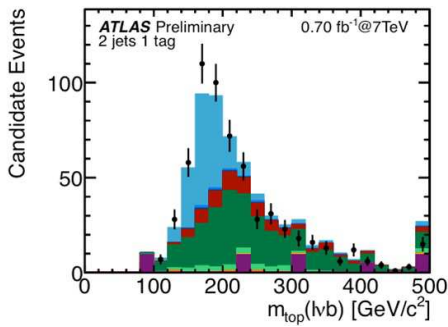} }
\caption{Lepton-neutrino-bjet mass spectrum in ATLAS.}
\label{fig:15}
\end{figure}

Production in the s-channel, and associated t-W production were also searched for by both experiments. See\linebreak P. Haefner's presentation at this Conference.

\subsection{Di-boson production}
\label{subsec:3-5}

Di-boson production provides stringent tests of the standard model (measurement of triple-gauge boson couplings), and represents at the same time benchmark reactions to assess several important modes for the Higgs boson search.

Most representative of both of these aspects is the ZZ production. Results obtained by ATLAS in the 4-lepton channel \cite{Ref17} are shown below as an example, CMS having similar performances. Events are triggered by either an electron or a muon of high transverse momentum. The analysis requires 4 leptons of p$_T >$15 GeV. At least two pairs of opposite charge need to fall in the Z mass window (66$<M_{l^+l^-}<$116 GeV). In the first 1.02 fb$^{-1}$ of 2011 data, 12 events were observed (2/4e, 8/4$\mu$,and 2/e$\mu$) while 0.3 events were expected from background (see Fig. 16). The corresponding fiducial cross-section was extracted to be \cite{Ref17}

\begin{equation}
\sigma = 19^{+6}_{-5}\ (stat.) \pm 1\ (syst.) \pm 1\ (lumi.)\ fb
\end{equation}

\noindent
and the channel cross-section:

\begin{equation}
\sigma^{ZZ}_{ATLAS} =  8.5^{+2.7}_{-2.3}\ (stat.)
\ ^{+0.4}_{-0.3}\ (syst.) \pm 0.3\ (lumi.)\ pb
\end{equation}

\noindent
to be compared with \cite{Ref18}

\begin{equation}
\sigma^{ZZ}_{CMS}  = 3.8 ^{+1.5}_{-1.2}\ (stat.)
 \pm 0.2\ (syst.) \pm 0.2\ ( lumi.)\ pb
\end{equation}

\noindent
and 6.5$^{+0.3}_{-0.2}$ pb  from NLO predictions.

\begin{figure}[h]
\vspace*{-0,2cm}\centering
\resizebox{0.70\columnwidth}{!}{%
  \includegraphics{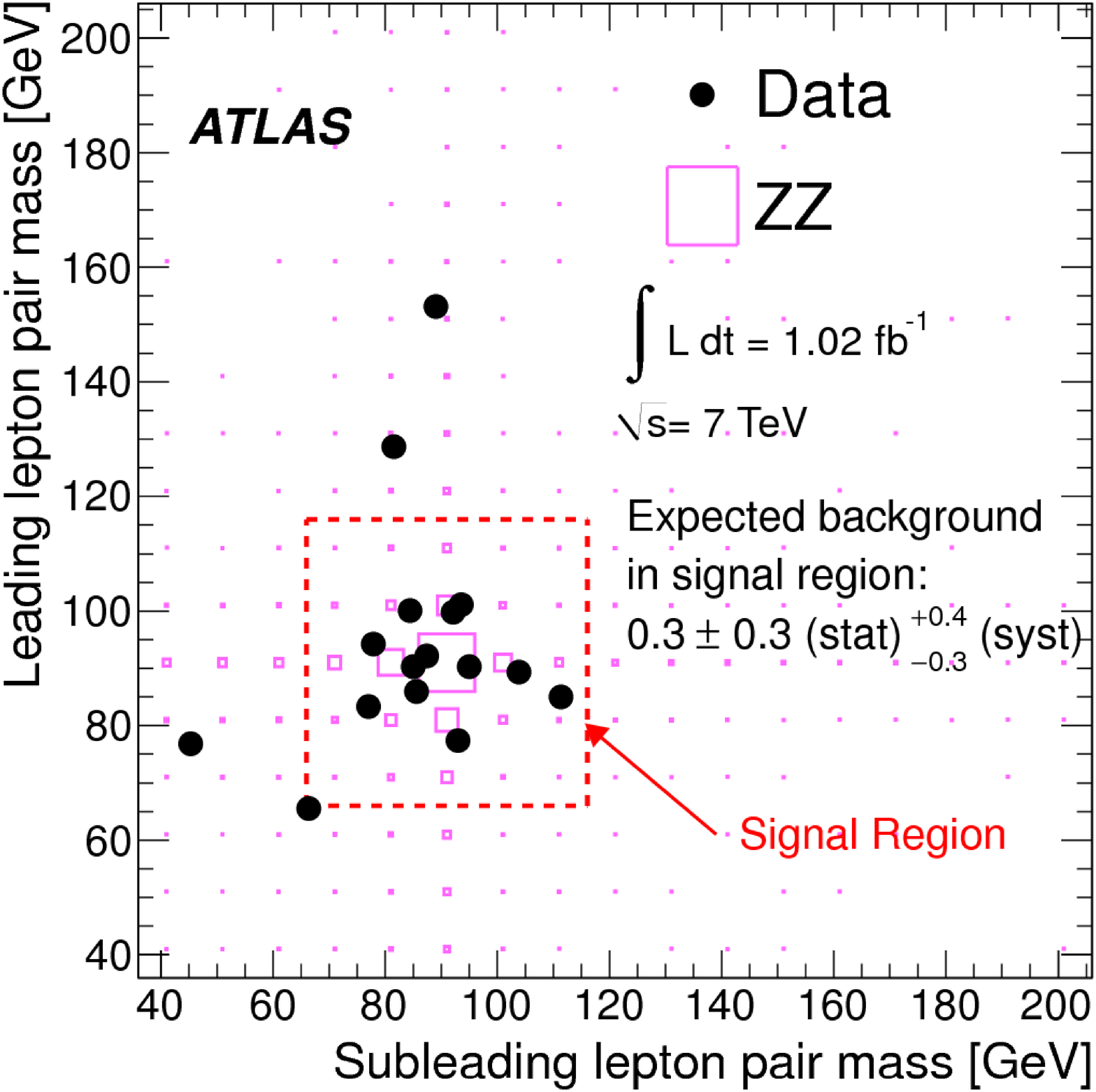} }
\caption{Lepton pairs invariant masses for  ZZ events in ATLAS.}
\label{fig:16}
\end{figure}

Given that the ZZZ and the ZZ$\gamma$ couplings are forbidden in the standard model, ATLAS extracted from the cross-section measurement the best limit todate on the corresponding f4 and f5 anomalous couplings \cite{Ref17}.

As a summary of standard model analyses already made by ATLAS and CMS, Fig. 17 shows a comparison of measured and predicted cross-sections in the case of CMS. The figure also includes the information concerning Vector bosons + N jets, not discussed here.

\section{Higgs search: Status and forecast}
\label{sec:3}

One of the events at this Conference is the presentation (see talk by L. Rolandi) of the combined search for the Higgs boson by the two collaborations, with up to about 2 fb$^{-1}$ for each of them. The individual results had already been presented before, and are summarized below:
\vspace*{0,1cm}

\noindent - ATLAS  excludes at 95\% CL (CLs limits) that the Standard Model Higgs boson be between 145 and 466 GeV, with the exception of two narrow bands (232 to 256 and 282 to 296 GeV). See Fig. 18 for the expected (131-447 GeV in the absence of a SM Higgs signal) and the observed limits.

\begin{figure}[h!]
\vspace*{-0,3cm}
\centering
\resizebox{0.80\columnwidth}{!}{%
  \includegraphics{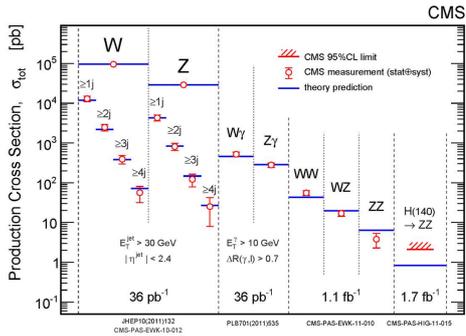} }
\caption{Standard model cross-sections in CMS.}
\label{fig:17}
\end{figure}

\begin{figure}[h]
\vspace*{-0,5cm}\centering
\resizebox{0.72\columnwidth}{!}{%
  \includegraphics{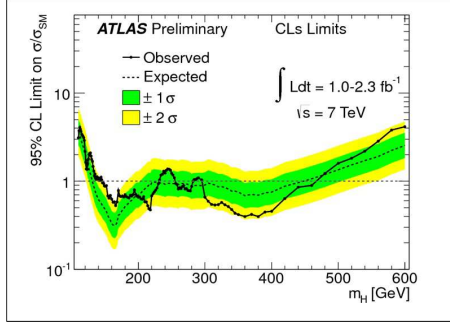}}
\caption{95\% CL upper limits for ``combined'' Higgs searches in ATLAS.}
\label{fig:18}
\end{figure}

\noindent
- CMS   excludes at 95\% CL that the SM Higgs boson be between 145 and 400 GeV, with the exception of two narrow bands, different from the ATLAS ones, (216 to 226 GeV and 288 to 310 GeV). See Fig. 19 for the expected limits (130-440 GeV) and the observed ones.

\begin{figure}[h]
\vspace*{-0,2cm}\centering
\resizebox{0.80\columnwidth}{!}{%
 \includegraphics{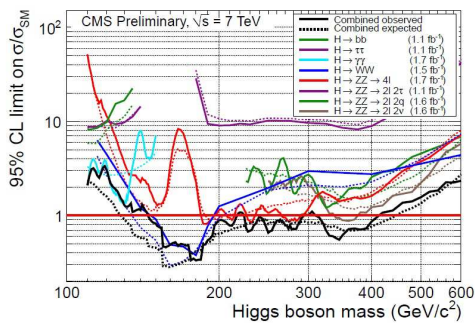} }
\caption{95\% CL upper limits  for ``combined'' Higgs searches in CMS.}
\label{fig:19}
\vspace*{-0,3cm}
\end{figure}

The main message from these two results is that the best motivated low mass region (the EW fits give\linebreak m$_H <$161 GeV at 95\% CL) is still open to exploration, while a wide ``medium/high'' mass range is excluded. 

While the high mass range (above $\sim$ 450 GeV) should not be prematurely discarded, it is clear that the largest effort in the short term will be devoted to the low mass region (m$_H <$ 145 GeV), where the main channels are\linebreak $H \to \gamma\gamma$, $H\to \tau\tau$, $H\to ZZ^* \to 4l$, $H \to WW^* \to ll\nu\nu$ and $VH,\ H\to b\bar b, V \to ll, l\nu$ or $\nu\nu$.

Given the extreme importance of these channels, and in particular of the first three which can give rise to a narrow mass peak, a short review of the expected performance of the experiments is given below. 

\subsection{Two-photon final state in ATLAS}
\label{subsec:4-1}

The huge background from jet-jet and $\gamma$-jet final states is mostly rejected by shower-shape cuts which take advantage of the high granularity of the ``accordion'' liquid argon electromagnetic calorimeter, featuring in particular three samplings in depth, and narrow strips ($\delta\eta$=0.008 $\times$ $\delta\phi$=0.1) in the first sampling which provide additional rejection against jets fragmenting with a leading $\pi^0$. An additional handle is provided by calorimetric photon isolation (a typical cut is a transverse energy cut of 5 GeV in a cone of $\Delta$R=0.4). Isolation provides a way to estimate the purity of the selected sample, found to be (for m$_{\gamma\gamma}>$ 100 GeV) $\sim$75\% prompt $\gamma\gamma$ and $\sim$25\% $\gamma$-jets, with much smaller contributions of jet-jet and Drell-Yan $e^+e^-$ pairs.

Thanks to the samplings in depth, electromagnetic calo\-rimeter data alone allow to measure the polar angle of each photon, and the space angle $\theta$ between the two photons. The accuracy of this measurement is illustrated  by the difference in the longitudinal position of the primary vertex found by intersecting the beam line  successively by each of the two photon's direction. This difference has an rms of 30mm, thus corresponding  to a primary vertex accuracy of 15mm, well below the longitudinal spread of primary vertices (about 56 mm), and accurate enough to give a contribution to the m$_{\gamma\gamma}$  resolution negligible compared to the effect of the photon energies resolution. When one or both of the photons are converted in the inner detector volume, the coordinate of this conversion point is used, with the shower barycentre, to give an even more accurate photon direction.

\begin{figure}[h]
\vspace*{-0,2cm}
\centering
\resizebox{0.75\columnwidth}{!}{%
  \includegraphics{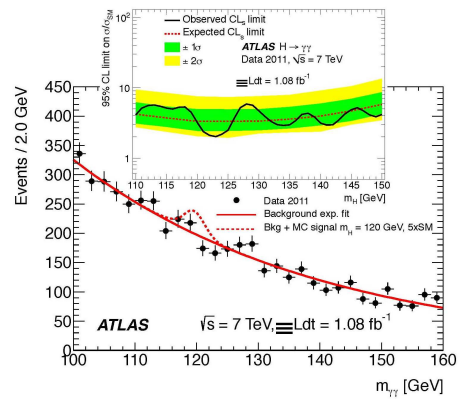} }
\caption{ATLAS $\gamma\gamma$ mass spectrum and associated limit.}
\label{fig:20}
\end{figure}

The energy response of the calorimeter is calibrated using Z$^0$ decays in $e^+e^-$ pairs. Monte-Carlo simulations are used to take into account the small differences in\linebreak response between electrons, converted photons, and unconverted photons. With the available statistics, the calibration was made by bins of $\delta\eta$=0.1, without any subdivision in azimuth. By comparing the width of the Z$^0$ line shape to Monte-Carlo simulations, this procedure also allows to estimate the ``constant term'' of the energy\linebreak resolution, which, for the data set considered for this optimization (2010 data worth 36 pb$^{-1}$) was 1.1 $\pm$0.5\% in the barrel ($\vert \eta\vert <$ 1.37)  and 1.8 $\pm$ 0.6\% in the end-caps (1.52 $<\vert \eta\vert <$ 2.37). The estimated average $\gamma\gamma$ mass resolution is about 1.7 GeV for m$_{\gamma\gamma}$ = 130 GeV.

The $\gamma\gamma$ spectrum obtained with 1.08 fb$^{-1}$ of data is shown in fig. 20. The 95\% CL limit, normalized to the SM Higgs cross-section, times the branching ratio to the two-photon final state, is given in a small insert, showing that with this amount of data the experiment was sensitive to about  4 times the SM Higgs cross-section.

The main effort on ATLAS performances related to this channel is to improve the constant term of the energy resolution, particularly in the end-caps. With 100 times more data than used in the current Z$^0$ lineshape optimization, it is hoped that the nominal 0.7\% constant term will be reached. The effect of higher pile-up (in particular on isolation-see section 3.2) also needs to be assessed. 

The use of additional variables (transverse momentum of the $\gamma\gamma$ system, presence of additional jets, decay angles in the $\gamma\gamma$ system) are also being investigated.

\subsection{Two-photon final state in CMS}
\label{subsec:4-2}

The absence of longitudinal segmentation in the PbWO4 crystal calorimeter of CMS imposes to combine the interaction vertex position with the shower positions in the calorimeter to determine the space angle between the two photons. The vertex  is selected on the basis of the sum of the p$_T^2$  of the tracks associated to each reconstructed vertex, combined with the p$_T$  balance between the tracks and the two-photon system. For $<\mu>$=6.5 corresponding to the analyzed data set,it was estimated by Monte-Carlo simulation that in 83\% of the cases, the selected vertex is within 10mm of the true vertex, a distance small enough to give a negligible contribution to the invariant mass resolution. Photon identification is based on shower size and isolation. As opposed to ATLAS, the sum of tracks p$_T$ in the isolation cone ($\Delta$R=0.3) is used on its own, and combined with the transverse energy in the electromagnetic calorimeter as discriminating variable.The cut values are adjusted to give the best S/B ratio for a particular signal photon efficiency. In the end, the sample purity is similar to ATLAS.

In order to eliminate crystal transparency variations as a function of the luminosity  integrated in the preceding few hours/days, correction factors are determined from the crystal response to laser pulses distributed over the calorimeter during part of the cycle without collisions (``abort gap'' in particular) \cite{Ref21}. The corrections range was up to about 10\% for the analysed data set. As in ATLAS, the Z$^0$ line shape is used to set the energy scale, improve the calibration, and determine the constant term.The resolution is then transported by Monte-Carlo to the two-photon final state. It ranges from less than 1.5 GeV for barrel-barrel events to about 3 GeV for barrel-end cap events.

The resulting $\gamma\gamma$ spectrum is shown in Fig. 21. The 95\% CL limit, normalized to the SM Higgs cross-section, times the branching ratio to the $\gamma\gamma$ final state is given in a small insert, showing that with 1.70 fb$^{-1}$ of data the experiment was sensitive to about  3 times the SM Higgs cross-section. In terms of detector performance, CMS is making a big effort to reach the nominal constant term which had been set to 0.5\%.

Comparing the two experiments one can see that, combining efficiency, background rejection, accuracy in energy and angular measurements, their sensitivities are at present quite similar, as illustrated by the expected CLs limits, account taken of the relative amount of data analyzed by each experiment at the time of the Conference.

\begin{figure}[h]
\centering
\resizebox{0.85\columnwidth}{!}{%
  \includegraphics{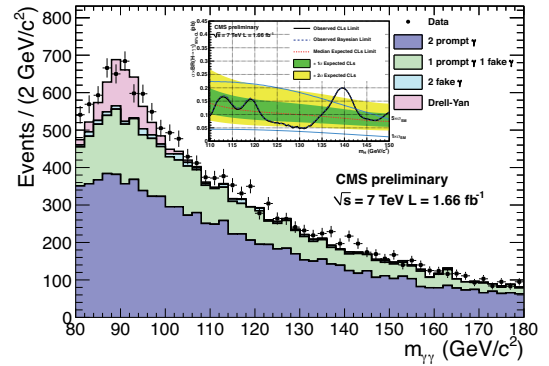} }
\caption{CMS $\gamma\gamma$ mass spectrum and associated limit.}
\label{fig:21}
\vspace{-0,8cm}
\end{figure}

\subsection{Four-lepton final state in ATLAS}
\label{subsec:4-3}

The $ZZ^* \to$ 4l final state combines low background and precision mass reconstruction, the main drawback being the small branching ratio. Non resonant ZZ production as considered in section 2.5 is an irreducible background, while $Z b\bar b$ and $t \bar t$ are the main other backgrounds, reduced by isolation and impact parameter requirements. The key point when addressing the low Higgs boson mass range is the reconstruction and identification efficiencies of leptons, and in particular electrons of transverse momenta down to 5 to 7 GeV. In an extended sample as compared to section 3.5 (up to 2.28 fb$^{-1}$), 27 events were selected (6ee, 9e$\mu$ and 12$\mu\mu$) with one Z mass requirement (76 to 106 GeV) while 28$\pm$4 were expected. One event only had a mass below 140 GeV, as shown in Fig. 22.
 
\begin{figure}[h!]
\centering
\resizebox{0.75\columnwidth}{!}{%
  \includegraphics{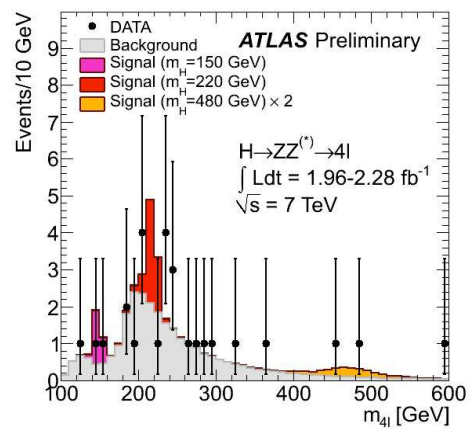} }
\caption{ATLAS 4l mass spectrum.}
\label{fig:22}
\end{figure}

The main performance effort for this channel, both in ATLAS and CMS, is towards improving the efficiency of soft leptons. On the analysis point of view the very low mass range (below 140 GeV) may benefit from considering events where the two Z are off mass shell, an attempt already made by CMS (see talks at this Conference).

\subsection{Higgs $\to \tau\tau$ in CMS}
\label{subsec:4-4}

This decay mode is a very useful complement to $\gamma\gamma$ in the low mass range, and a key channel for the MSSM Higgs for a large range of M$_A$ and tan$\beta$ values. Recently, CMS made public \cite{Ref22} an analysis with 1.6 fb$^{-1}$ of data, using the e-$\mu$, e-had and $\mu$-had final states, triggered by either a lepton or a lepton plus a $\tau$-jet (see section 3.3). The $\tau$-jets are reconstructed using particle flow and identified as briefly described in 3.3. One of the main backgrounds is W+jets, where the W decay leptonically, and the jet is misidentified as a $\tau$-jet. This background is rejected by cutting on the E$_T^{miss}$ projected onto the bisector of the two $\tau$ ``visible'' decay products. The analysis is made using the invariant mass of the two $\tau$  visible decay products. 

The sensitivity to the SM Higgs is enhanced by treating separately events with 2 additional jets in the final state, in a configuration compatible with Higgs production by Vector Boson Fusion ($\vert \Delta\eta_{jj}\vert >$ 3.5 and Mjj $>$ 350 GeV). No significant excess over background is seen, allowing to put a 95\% CL limit at $\sim$ 10 times the SM Higgs cross-section for M$_H$ = 130 GeV, while the expected value ranged from around 7 at 110 GeV up to about 9 at 140 GeV.

The sensitivity to the MSSM is enhanced by treating separately events with at least one b-tagged jet. No excess over background is seen, allowing to rule out,  at 95\% CL, a large fraction of the MSSM space, as shown in Fig. 23. It is interesting to note that for M$_A <$ 130 GeV the whole tan$\beta$ range is already ruled out. The superiority of LHC over the Tevatron in this channel is also exemplified by comparing the CMS limit to the D0 limit obtained with a more than four times larger data set.

\begin{figure}[h]
\centering
\resizebox{0.75\columnwidth}{!}{%
  \includegraphics{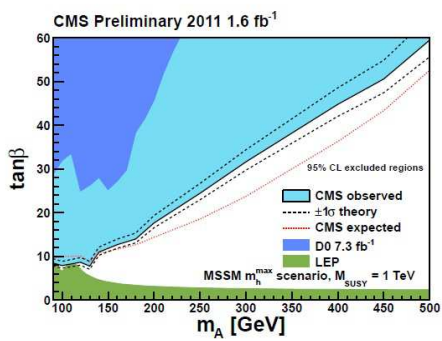} }
\caption{Excluded MSSM parameter space in CMS, using the $\tau-\tau$ channel.}
\label{fig:23}
\end{figure}

The main effort for the $\tau\tau$ final state, both in ATLAS (no results shown here) and CMS  will be to maintain the performance with values of $<\mu>$  reaching 20 and above.

\section{Search for new Physics}
\label{sec:5}

The search for excited quark states, axigluons, and W' was already mentioned in section 3.1. Only two examples of searches for SUSY effects, either direct or indirect are briefly described below (beyond the MSSM Higgs considered in 4.4) to illustrate the present situation, and the status of the search for Z' in lepton pairs is given. In all cases, for more recent updates, see dedicated talks at the Conference.

\subsection{B$_s \to \mu\mu$}
\label{subsec:5-1}

In the Standard Model this channel, which has similarities with the historically famous K$_L \to \mu\mu$ decay, is predicted to have a branching ratio of (3.2 $\pm$0.2) 10$^{-9}$. If supersymmetry is realized as in the MSSM, with large enough tan$\beta$ values, the branching ratio can significantly be  increased by the contribution of additional particles in the loops (see Fig. 24) which goes like tan$\beta^6$.

\begin{figure}[h]
\centering
\resizebox{0.75\columnwidth}{!}{%
  \includegraphics{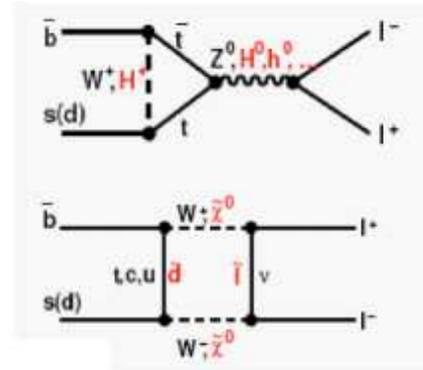} }
\caption{SUSY diagrams contributing to B$_s \to \mu\mu$.}
\label{fig:24}
\end{figure}

In CMS the analysis requires two muons of p$_T$ larger than 4 GeV, isolated and with a highly significant flight path (L/$\sigma >$ 15 (barrel) or $>$20 (End-Caps)). Events are counted in a window of $\pm$ 75 MeV around the nominal B$_s$ mass. The event count was found compatible with background only, leading to an upper limit on the branching ratio of 1.9 10$^{-8}$ at the 95\% CL, for a data set of 1.14 fb$^{-1}$ \cite{Ref23}. A similar analysis in LHCb \cite{Ref24} led to a limit of 1.5 10$^{-8}$. With the addition of the data sets not yet analyzed, and improved analysis, the standard model limit is becoming a target soon within reach. 

\subsection{Search for s-quarks and gluinos}
\label{subsec:5-2}

Search for supersymmetric particles was one of the priority topics when data at the LHC became available. In the R-parity conserving scenarios, standard searches require a large E$_T^{miss}$ together with several jets. Different channels can be addressed depending on the presence of one or more leptons in the final state, of the same sign, or of opposite sign.

A summary of the limits obtained by CMS \cite{Ref25} with up to 1.1 fb$^{-1}$ of data, is shown in Fig. 25. One observes that the more stringent limits are given by the ``Jets+ E$_T^{miss}$'' channel, for which the limit is slightly above 1 TeV for both squarks and gluinos, if one assumes they have similar masses. The limit by ATLAS (1.07 TeV) is similar. The gain of sensitivity with respect to the Tevatron is striking. Future searches will address more exclusive final states, like s-tops,... until the energy of the LHC is significantly increased.

\begin{figure}[h]
\centering
\resizebox{0.75\columnwidth}{!}{%
  \includegraphics{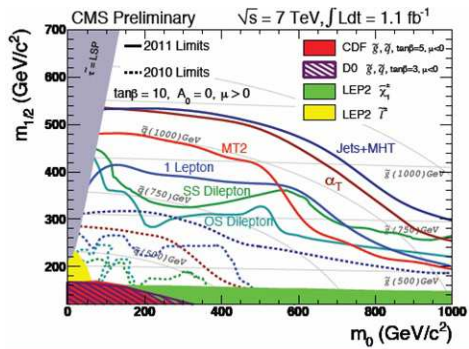} }
\caption{CMS limits for supersymmetry.}
\label{fig:25}
\end{figure}

\subsection{Search for heavy vector bosons (Z')}
\label{subsec:5-3}

Dilepton final states (ee or $\mu\mu$)  provide easy trigger and clean signatures for the search of heavy recurrences of the Z$^0$. The Sequential Standard Model Z' (same couplings as the Z$^0$) is most often taken as bench mark. In the muon final state the best possible alignment of precision chambers is required to maintain good accuracy in the upper invariant mass range. The results obtained by ATLAS with up to 1.08 fb$^{-1}$ are shown in Fig. 26. The reach in the framework of the SSM goes up to 1.83 TeV. One observes that, already with this luminosity, the limits are better than at the Tevatron (with $\sim$ 5 fb$^{-1}$) even in the ``low'' mass range (200 GeV and above). The limits obtained by CMS are similar. Final states with one lepton and large E$_T^{miss}$ are used to set limits on the W'. Significant further progress in these searches will have to wait for increased LHC centre of mass energy.

\begin{figure}[h]
\centering
\resizebox{0.80\columnwidth}{!}{%
  \includegraphics{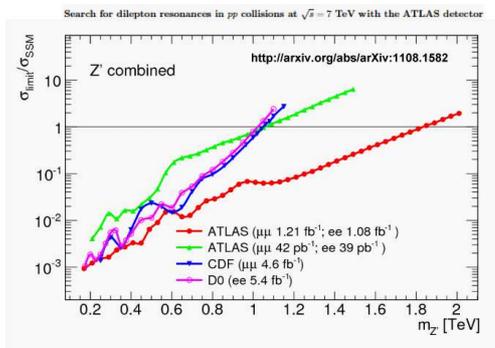} }
\caption{ATLAS limits for a sequential Z'.}
\label{fig:26}
\end{figure}

\section{Conclusions}

After some difficult times, the LHC has been running steadily in 2011, at 7 TeV in the centre of mass, with instantaneous luminosities reaching peak values above  3.5 10$^{33}$ cm$^{-2}$ s$^{-1}$. With such instantaneous luminosities and 50 ns bunch spacing, the average number of collisions per crossing is over 12. The ATLAS and CMS detectors have demonstrated excellent performance (including trigger and event reconstruction) in these already difficult conditions.

Analyses of ElectroWeak bench mark channels show very good agreement between data and the Standard Model. With the accumulated data sets, precision physics (M$_{top}$, M$_W$,...) can now start.

No phenomenon ``beyond Standard Model'' has so far shown-up,but this exploration is only beginning.

The hot-topic of 2011 and 2012 is the search for the SM Higgs boson. A very broad mass range has been excluded, but the best motivated low mass region (114 to 140 GeV) is still fully open. Getting the best out of present (and 2012) data, in this difficult mass range, requires pushing the precision of ``objects'' reconstruction, and the physics analyses, to their maximum.

We look forward hearing about updates, new ideas,... during the Conference.


\begin{thebibliography}{}
\bibitem{Ref1} The LHC machine.  JINST3 (2008) S08001, L. Evans \& P. Bryant editors
\bibitem{Ref2} P. Collier, CERN-LHCC meeting, Sept. 21st, 2011\\
 see: http://indico.cern.ch/getFile.py/access?contribId=1\&\linebreak sessionId=0\&resId=1\&materialId=slides\&confId=153317
\bibitem{Ref3} ATLAS Collaboration, Luminosity Determination\\ in pp Collisions at $\sqrt{s}$=7 TeV using the ATLAS Detector at the LHC. EPJC 71(2011) 1630
\bibitem{Ref4} CMS Collaboration: Absolute Calibration of the CMS Luminosity Measurement:Summer 2011. Update CMS-PAS-EWK-11-001
\bibitem{Ref5} Beam induced backgrounds in the ATLAS experiment. D. Salek, Poster presented at this Conference
\bibitem{Ref6} See the talk by K. Eggert at this Conference
\bibitem{Ref7} ATLAS Collaboration, The ATLAS experiment at the CERN LHC. JINST 3(2008) S08003
\bibitem{Ref8} CMS Collaboration, The CMS experiment at the CERN LHC. JINST 3 (2008) S08004
\bibitem{Ref9} ATLAS Collaboration. See\\
 https://twiki.cern.ch/twiki/bin/view/AtlasPublic/\linebreak StandardModelPublicCollisionPlots
\bibitem{Ref10} ATLAS Collaboration, Measurement of the inclusive W$^{\pm}$ and Z/$\gamma$ cross sections in the electron and muon decay channels in pp collisions at $ \sqrt{s}$ = 7 TeV with the ATLAS detector. arXiv 1109.5141 submitted to\linebreak Phys. Rev. D
\bibitem{Ref11} CMS Collaboration, Measurement of the Inclusive Z Cross Section via Decays to Tau Pairs in pp Collisions at $\sqrt{s}$ = 7 TeV.  JHEP 1108 :117 (2011)
\bibitem{Ref12} ATLAS Collaboration,  Commissioning of the ATLAS high-performance b-tagging algorithms in the 7 TeV collision data. ATL-CONF-2011-102
\bibitem{Ref13} ATLAS Collaboration: Measurement of the top quark pair production cross-section based on a statistical combination of measurements of dilepton and single-lepton final states at $\sqrt{s}$ = 7 TeV with the ATLAS detector. ATL CONF-2011-108
\bibitem{Ref14} CMS Collaboration, Combination of top pair production cross section measurements. CMS-PAS-TOP-11-024
\bibitem{Ref15} CMS Collaboration, Measurement of the mass difference between top and antitop quarks. CMS-PAS-TOP-11-019
\bibitem{Ref16} Measurement of the t-channel Single Top-Quark Production Cross-section in 0.70 fb$^{-1}$ of pp Collisions at $\sqrt s$ = 7 TeV collected with the ATLAS detector. ATL-CONF-2011-101
\bibitem{Ref17} Measurement of the ZZ production cross section and limits on anomalous neutral triple gauge couplings in proton-proton collisions at $\sqrt  s$ = 7 TeV with the ATLAS detector. Accepted by Phys. Rev. Lett.
\bibitem{Ref18} Measurement of the WW, WZ and ZZ cross sections at CMS. CMS-PAS-EWK-11-010
\bibitem{Ref19} ATLAS Collaboration, Update of the Combination of Higgs Boson Searches in 1.0 to 2.3 fb$^{-1}$ of pp Collisions data taken at $\sqrt{s}$ = 7 TeV with the ATLAS Experiment at the LHC. ATL-CONF-2011-135
\bibitem{Ref20} CMS Collaboration, Combined Standard Model Higgs boson searches with up to 2.3 inverse femtobarns of pp collision data at $\sqrt{s}$=7 TeV at the LHC. CMS-PAS-HIG-11-023
\bibitem{Ref21} Anfreville M. et al., Laser monitoring system for the CMS lead tungstate crystal calorimeter. CMS-NOTE-2007-028 
\bibitem{Ref22} CMS Collaboration, Search for Neutral Higgs Bosons decaying to Tau Pairs in pp Collisions at $\sqrt{s}$=7 TeV. CMS PAS HIG-11-020
\bibitem{Ref23} Search for B$^0_s \to \mu^+\mu^-$ and B$^0 \to \mu^+\mu^-$ decays in pp collisions at $\sqrt s$ = 7 TeV. The CMS Collaboration.\break Phys. Rev. lett. 107, 191802 (2011)
\bibitem{Ref24}LHCb Collaboration, Search for the rare decays B$^0_s \to \mu^+ \mu^-$ and B$^0 \to \mu^+ \mu^-$,  arXiv hep-ex/1112.1600 submitted to Phys. Lett. B
\bibitem{Ref25} CMS Collaboration, Search for Supersymmetry at the LHC in Events with Jets and Missing Transverse Energy. Phys. Rev. Lett. 107, 221804 (2011)
\bibitem{Ref26} ATLAS Collaboration, Search for dilepton resonances in pp collisions at $\sqrt{s}$ = 7 TeV with the ATLAS detector. Phys. Rev. Lett. 107.272002 (2011)


\end{thebibliography}
\end{document}